\documentclass[a4paper,10pt]{article}
\usepackage{graphicx}

\begin{document}

\title{Phase transition of two-dimensional Ising models
on the honeycomb and related lattices
with striped random impurities
}

\author{Satoshi Morita$^1$ and Sei Suzuki$^2$
\vspace*{5pt}\\
{\itshape $^1$Institute for Solid State Physics, University of Tokyo,}\\
{\itshape Kashiwa, Chiba 277-8581, Japan}\\
{\itshape $^2$Department of Liberal Arts, Saitama Medical University,}\\
{\itshape Moroyama, Sitama 350-0495, Japan}}

\date{}

\maketitle

\begin{abstract}
Two-dimensional Ising models on the honeycomb lattice and the square
lattice with striped random impurities are studied to obtain their phase
diagrams.
Assuming bimodal distributions of the random impurities where all the
non-zero couplings have the same magnitude,
exact critical values for the fraction
$p$ of ferromagnetic bonds at the zero-temperature
($T=0$) are obtained.
The critical lines in the $p-T$ plane are drawn by
numerically evaluating the Lyapunov exponent of random matrix
products.
\end{abstract}

\section{Introduction}
\label{intro}

Ising models have attracted a particular attention
in the statistical physics as simplest models that
exhibit phase transitions and critical phenomena.
The most popular is the pure Ising ferromagnet model on the square
lattice, for which
not only the transition temperature \cite{bib:Kramers1941}
but also the free energy have been obtained exactly \cite{bib:Kaufman1949,bib:Onsager1944}.
The hexagonal (honeycomb) lattice is the secondly simplest
two-dimensional lattice.
The pure Ising model on the honeycomb lattice has been studied by
Wannier \cite{bib:Wannier1950} and Houtapel \cite{bib:Houtappel1950} and exactly solved.
Apart from the Ising model, quantum spin models such as
the Kitaev model \cite{bib:Kitaev2006} or
Kitaev-Heisenberg model \cite{bib:Chaloupka2010,bib:Jackeli2009}
on the honeycomb lattice have recently received
a lot of theoretical and experimental attentions
\cite{bib:Chaloupka2013,bib:Choi2012,bib:Reuther2011,bib:Singh2012,bib:Ye2012}.
Such a growing interest in the honeycomb-lattice systems
motivates us to revisit the Ising model on the
honeycomb lattice.

In contrast to pure models,
less is understood for disordered Ising models.
The exact free energy has not been obtained in fully
random Ising models in two dimension.
Although the transition temperature and/or phase diagram
have been investigated using the replica trick, few is
revealed so far.
Domany showed for the diluted Ising model
on the square lattice with concentration $p$ of the ferromagnetic bonds
that the zero-temperature phase transition occurs
at $p_c = 1/2$ \cite{bib:Domany1978}.
Fisch \cite{bib:Fisch1978}, Schwartz \cite{bib:Schwartz1979},
and Aharony and Stephen \cite{bib:Aharony1980}
obtained equations to determine the
transition temperature for square-lattice models with
random interactions each of which takes $J_1$ or $J_2$
with the same probability.

\begin{figure}[t]
\begin{center}
\includegraphics[scale=0.5]{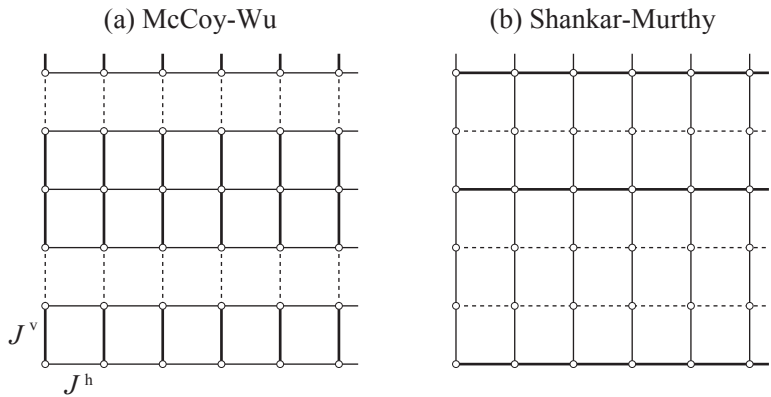}
\end{center}
\caption{Schematics for (a) the McCoy-Wu model and (b) the Shankar-Murthy model.
(a) In the McCoy-Wu model, the interactions $J_{ij}^{\rm h}$'s
on the horizontal bonds are uniform and $J_{ij}^{\rm v}$'s 
on the vertical bonds are random from row to row but uniform
from column to column. (b) In the Shankar-Murthy model,
$J_{ij}^{\rm h}$'s are random from row to row but uniform among
a row, while $J_{ij}^{\rm v}$'s are entirely uniform.}
\label{fig:Lattice}
\end{figure}

The problem is simpler if the disorder or impurity exists only in one
direction and the system maintains the translational invariance
in the other direction.
Let us consider an Ising model on the square lattice.
Suppose that $J_{ij}^{\rm v}$ and $J_{ij}^{\rm h}$
denote coupling constants on a vertical and a horizontal
bonds respectively.
McCoy and Wu \cite{bib:McCoy1968} introduced a model where
$J_{ij}^{\rm h}$'s are constant and
$J_{ij}^{\rm v}$'s are uniform among the columns
but row-to-row random. See Fig. \ref{fig:Lattice}.
Shankar and Murthy \cite{bib:Shankar1987} considered a slightly different model,
where $J_{ij}^{\rm h}$'s are uniform among a row
but row-to-row random while $J_{ij}^{\rm v}$'s are uniform
in the whole system (Fig. \ref{fig:Lattice}).
The criticality condition for both the models was derived in
Refs. \cite{bib:Wolff1981,bib:Kardar1982}.
It is important that the partition function of the McCoy-Wu and the
Shankar-Murthy models can be written as the largest eigenvalue, or
the Lyapunov exponent in other words,
of a product of random $2\times 2$ matrices.
As far as the transition temperature is concerned, however,
it is sufficient to work with random diagonal matrices.

In the present paper, we consider random Ising models with striped
randomness on the honeycomb and square lattices.  These models are
extensions of the Shankar-Murthy model, and were discussed earlier by
Hamm \cite{bib:Hamm1977} and Hoever \cite{bib:Hoever1982}.  The random
couplings are not always ferromagnetic.  As we shall see below, in these
models one encounters computation of the Lyapunov exponent of a product
of random non-diagonal matrices to determine the location of a phase
transition.  This is contrasted with the situation for similar but
inequivalent models with ferromagnetic couplings discussed in
Refs. \cite{bib:Igloi1995}, for which the transition temperature is
determined exactly by a single numerical equation.  Unfortunately, there
is no complete mathematical method to compute such a Lyapunov exponent.
In the present paper, nevertheless, we successfully compute the exact
Lyapunov exponent in the limiting case, where all the non-zero couplings
have the same magnitude and the temperature is zero, and provide the
exact location of the zero-temperature phase transition on the axis of
the density of impurities.  We furthermore provide precise phase
diagrams on the basis of the numerical computation of Lyapunov
exponents.

The outline of the present paper is as follows.  We first define the
models in Sect. \ref{sec:Model}.  We then introduce the transfer
matrices and their Majorana-fermion representations in
Sect. \ref{sec:Trans}.  We derive an equation which determines the
transition temperature there.  On the basis of the equation derived in
Sect. \ref{sec:Trans}, we visit known exact results for pure systems in
Sect. \ref{sec:Pure}.  We then give main results on random systems in
Sect. \ref{sec:Random}, where the zero-temperature phase transition is
first discussed and finite temperatures follow it.  Section
\ref{sec:Concl} is devoted to the conclusion.

\section{Models}
\label{sec:Model}

\begin{figure}[t]
\begin{center}
\includegraphics[scale=0.4]{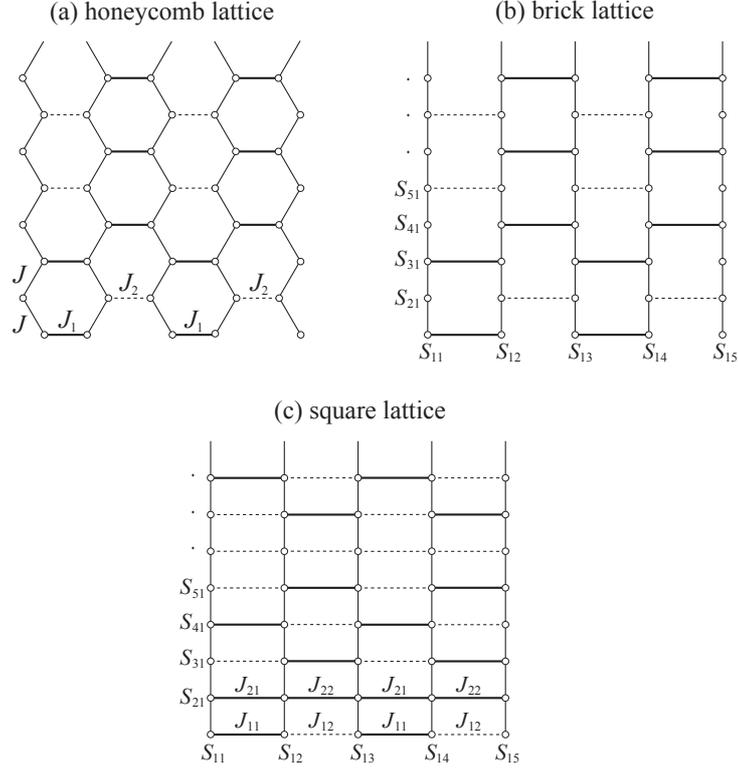}
\end{center}
\caption{Lattices studied in the present paper.
(a) The honeycomb lattice, (b) the brick lattice, and
(c) the square lattice.
(a) and (b) are equivalent.
For the brick lattice (i.e., the honeycomb lattice), we assume that
the interactions on the vertical bonds denoted by $J$ are uniform,
while those on the horizontal bonds denoted by $J_i$
are uniform in a row but row-to-row random.
As for the square lattice,
we assume the same property of the interaction on the vertical
bonds. The interactions on the horizontal bonds, however,
are independently random in the adjoining two bonds
and alternate in a row, and besides they are row-to-row random.
}
\label{fig:Honeycomb}
\end{figure}

Let us begin with the honeycomb-lattice model. As shown in
Fig. \ref{fig:Honeycomb}, the honeycomb
lattice is equivalent to the brick lattice.
We write the Ising spin sitting on each lattice
point of the brick lattice
as $S_{i,j}$, where $i$ and $j$ label vertical and
horizontal positions respectively.
We then assign a spin-spin interaction $J$ on the
vertical bonds and $J_{i}$ on the horizontal
bonds, where $i$ labels the vertical position
of a bond. We assume that $J_{i}$'s are random and independent for
different $i$.
The Hamiltonian is written as
\begin{eqnarray}
 H &=& - J\sum_{i=1}^{M}\sum_{j=1}^{N} S_{i,j} S_{i+1,j}
  \label{eq:H_brick}
  \\
 && - \sum_{\mu=1}^{M/2}\sum_{\nu=1}^{N/2}
  \left(J_{2\mu - 1}S_{2\mu-1, 2\nu-1}S_{2\mu-1,2\nu}
  + J_{2\mu}S_{2\mu,2\nu}S_{2\mu,2\nu+1}\right) ,
  \nonumber
\end{eqnarray}
where $M$ and $N$ denote the number of sites
on the vertical and horizontal lines respectively.
We assume that $M$ and $N$ are even numbers.
We consider two bimodal distributions for
random $J_{i}$ as follows:
\begin{equation}
 \begin{array}{l@{\hspace{1em}}l}
  \mbox{($\pm J$ model)} &
   P(J_{i}) = p \,\delta (J_{i} - J_0)
   + (1 - p)\,\delta (J_{i} + J_0) ,\\
  \\
  \mbox{(diluted model)} &
   P(J_{i}) = p \,\delta (J_{i} - J_0)
   + (1 - p)\,\delta (J_{i})  , \\
 \end{array}
\label{eq:DistJ_ij}
\end{equation}
where $J_0 > 0$ and $0 \leq p \leq 1$. Note that $p$
stands for the density of ferromagnetic bonds in the
horizontal bonds. We moreover restrict ourselves to the
range of $p$ such that $[J_i] > 0$, where
the notation $[\cdots]$ means the average over
the randomness of the bonds:
\begin{equation}
 [\cdots] = \int \prod_i dJ_iP(J_i) (\cdots).
\end{equation}

We next consider the square lattice.
Same as the brick lattice, we assume that the interactions
on the vertical bonds are uniform. Regarding the horizontal
bonds, however, we assume that the interactions are
random in the adjoining two bonds and form an alternating
order, and besides they are independent in different rows.
An instance of the configuration of interactions is depicted in
Fig. \ref{fig:Honeycomb}(c). The Hamiltonian of the present
model is given by
\begin{eqnarray}
 H &=& - J\sum_{i=1}^{M}\sum_{j=1}^{N} S_{i, j} S_{i+1, j}
  \label{eq:H_square}\\
 && - \sum_{i=1}^{M}\sum_{\nu=1}^{N/2}
  \left(J_{i1}S_{i, 2\nu-1}S_{i,2\nu}
  + J_{i2}S_{i,2\nu}S_{i,2\nu+1}\right) ,
  \nonumber
\end{eqnarray}
where $J_{i1}$ and $J_{i2}$ ($i=1,2,\cdots, M$)
are independently random and follow Eq.~(\ref{eq:DistJ_ij}).

For both lattices, we assume the periodic boundary condition
in the horizontal direction ($S_{i, N+1} = S_{i,1}$)
and in the vertical direction ($S_{M+1, j} = S_{1,j}$).

It should be noted that the $\pm J$ models on both lattices
mentioned above are invariant under the change of the sign of all the
horizontal couplings and that of the spins in every odd
column.  Therefore it turns out that the $\pm J$ models have the
ferromagnetic-antiferromagnetic symmetry with respect to $p = 1/2$.

\section{Transfer matrices}
\label{sec:Trans}

We first focus on the square-lattice model.
A portion of the Boltzmann factor regarding the lowest two layers in
Fig.~\ref{fig:Honeycomb}(c) is written as
\begin{eqnarray}
 &&\exp\left(
      \sum_{\nu=1}^{N/2}\left(K_{11}S_{1,2\nu-1}S_{1,2\nu}
      + K_{12}S_{1,2\nu}S_{1,2\nu+1}\right)
      + L\sum_{j=1}^{N} S_{1,j}S_{2,j} \right.\nonumber\\
 &&+ \left.
      \sum_{\nu=1}^{N/2}\left(K_{21}S_{2,2\nu-1}S_{2,2\nu}
      + K_{22}S_{2,2\nu}S_{2,2\nu + 1}\right)
      + L\sum_{j=1}^{N} S_{2,j}S_{3,j}
\right),
\end{eqnarray}
where $K_{ij} = J_{ij}/T$ ($i=1,2,\cdots,M$; $j=1,2$) and $L = J/T$ with
the temperature $T$ in the unit of $k_B = 1$.
The transfer matrix between the lowest and the second lowest rows
is given by
\begin{eqnarray}
 &&T_{K_{11},K_{12},K_{21},K_{22}}
  (S_{11},\cdots,S_{1N};S_{31},\cdots,S_{3N})
  \nonumber\\
 &&=\sum_{S_{21},\cdots,S_{2N}}
  A_{K_{11},K_{12}}(S_{11},\cdots,S_{1N};S_{21},\cdots,S_{2N}) \nonumber\\
 &&\hspace{2cm}\times A_{K_{21},K_{22}}(S_{21},\cdots,S_{2N};S_{31},\cdots,S_{3N}) ,
\end{eqnarray}
where
\begin{eqnarray}
 &&A_{K_{11},K_{12}}(S_{11},\cdots,S_{1N};S_{21},\cdots,S_{2N})
  \nonumber\\
 &&= \exp\left(\sum_{\nu=1}^{N/2}\left(K_{11}S_{1,2\nu-1}S_{1,2\nu}
        + K_{12}S_{1,2\nu}S_{1,2\nu+1}\right)
  + L\sum_{j=1}^{N}S_{1j}S_{2j}\right) .
\label{eq:A}
\end{eqnarray}
Using these $2^{N}\times 2^{N}$ transfer matrices, the partition function
is written as
\begin{equation}
 Z = {\rm
  tr}\left(T_{K_{11},K_{12},K_{21},K_{22}}T_{K_{31},K_{32},K_{41},K_{42}}
      \cdots T_{K_{M-1,1},K_{M-1,2},K_{M1},K_{M2}}\right) .
\end{equation}
The matrices in Eqs.~(\ref{eq:A}) are now written as
\begin{equation}
 A_{K_{11},K_{12}}(S_{11},\cdots,S_{1N};S_{21},\cdots,S_{2N})
  = C\langle S_{11},\cdots,S_{1N}|
  \hat{A}_{K_{11},K_{12}}
  |S_{21},\cdots,S_{2N}\rangle
  \label{eq:A1}
\end{equation}
\begin{equation}
 \hat{A}_{K_{11},K_{12}} =
  \exp\left(\sum_{\nu=1}^{N/2}\left(K_{11}\sigma_{2\nu-1}^z\sigma_{2\nu}^z
    + K_{12}\sigma_{2\nu}^z\sigma_{2\nu+1}^z\right)\right)
  \exp\left(g\sum_{j=1}^{N}\sigma_j^x\right)
\end{equation}
where $\sigma_j^{\alpha}$ ($\alpha=x, y, z$) is the
$\alpha$-component of the Pauli operator
and $\{|S_{i1},\cdots,S_{iN}\rangle\}$ is the basis
of an $N$-spin system satisfying
$\sigma_j^z|S_{i1},\cdots,S_{iN}\rangle =
S_{ij}|S_{i1},\cdots,S_{iN}\rangle$ for $j=1,2,\cdots,N$.
$C$ and $g$ are defined by
\begin{equation}
 C = \left(\frac{1}{2}\sinh 2g\right)^{-N/2} ,~~~
  g = \frac{1}{2}\log\coth L .
\end{equation}
We introduce Majorana fermions and their Fourier transformations as
follows;
\begin{equation}
 \psi_1(\nu)=\frac{1}{\sqrt{2}}
  \left(\prod_{k=1}^{2\nu-2}\sigma_k^x\right)
  \sigma_{2\nu-1}^z, \quad
 \psi_2(\nu)=\frac{1}{\sqrt{2}}
  \left(\prod_{k=1}^{2\nu-2}\sigma_k^x\right)
  \sigma_{2\nu-1}^y,
\end{equation}
\begin{equation}
 \psi_3(\nu)=\frac{1}{\sqrt{2}}
  \left(\prod_{k=1}^{2\nu-1}\sigma_k^x\right)
  \sigma_{2\nu}^z, \quad
 \psi_4(\nu)=\frac{1}{\sqrt{2}}
  \left(\prod_{k=1}^{2\nu-1}\sigma_k^x\right)
  \sigma_{2\nu}^y,
\end{equation}
\begin{equation}
 \psi_n(\nu) = \sqrt{\frac{2}{N}}\sum_{0<q<\pi}
  \left( c_n(q)e^{iq\nu} + c_n^\dagger(q)e^{-iq\nu} \right),
\end{equation}
where $n=1,2,3,4$ and $\nu$ runs from $1$ to $N/2$.
Although the possible values of the wave number $q$ depend
on the parity of $\prod_{j=1}^N\sigma_j^x$, they
have no effect on the argument below and hence we do not
care about them.
Then $\hat{A}_{K_{11},K_{12}}$ in Eq.~(\ref{eq:A1})
is written as $\hat{A}_{K_{11},K_{12}} = \exp(H_{K_{11},K_{12}}^{(1)})
\exp(H_g^{(2)})$,
where
\begin{eqnarray}
 H_{K_{11},K_{12}}^{(1)} &=& 2i\sum_{\nu=1}^{N/2}
  \left(K_{11}\psi_2(\nu-1)\psi_3(\nu)
  + K_{12} \psi_4(\nu)\psi_1(\nu+1) \right)
  \nonumber\\
 &=& 2i\sum_{0<q<\pi}
  \left\{K_{11}
   \left(
    c_2(q)c_3^{\dagger}(q)
    + c_2^{\dagger}(q)c_3(q)
   \right)
  \right. \nonumber\\
 &&~~~~~~
  +\left. K_{12}
    \left(e^{-iq}c_4(q)c_1^{\dagger}(q)
     +e^{iq}c_4^{\dagger}(q)c_1(q)
    \right)
   \right\}
  \nonumber\\
 &\equiv& \sum_{0<q<\pi}h_{K_{11},K_{12}}^{(1)}(q) ,
\end{eqnarray}
\begin{eqnarray}
 H_g^{(2)} &=& 2i g\sum_{\nu=1}^{N/2}
  \left(\psi_1(\nu)\psi_2(\nu)
   +\psi_3(\nu)\psi_4(\nu)\right) \nonumber\\
 &=& 2i g\sum_{0<q<\pi}
  \left(
   c_1(q)c_2^{\dagger}(q)
   + c_1^{\dagger}(q)c_2(q)
   + c_3(q)c_4^{\dagger}(q)
   + c_3^{\dagger}(q)c_4(q)
  \right) \nonumber\\
 &\equiv& \sum_{0<q<\pi} h_g^{(2)}(q) .
\end{eqnarray}
Since each mode is decoupled from others, the transfer
matrices are given by a product of those with a fixed mode.
Hence one gets
\begin{equation}
 Z = \prod_{0<q<\pi}{\rm tr}\left\{
   \hat{\mathcal{A}}_{K_{11},K_{12}}(q)\hat{\mathcal{A}}_{K_{21},K_{22}}(q)\cdots
   \hat{\mathcal{A}}_{K_{M1},K_{M2}}(q) \right\} ,
\label{eq:Z1}
\end{equation}
where
\begin{equation}
 \hat{\mathcal{A}}_{K_{11},K_{12}}(q)
  = \exp(h_{K_{11},K_{12}}^{(1)}(q))\exp(h_{g}^{(2)}(q)) .
\end{equation}
Taking the limit $N\to\infty$, the free energy per spin is written as
\begin{equation}
 f = \int_{0}^{\pi}\frac{dq}{2\pi} f(q) ,
\end{equation}
where
\begin{equation}
 f(q) = - T\frac{1}{M}\ln{\rm tr}\left\{
    \hat{\mathcal{A}}_{K_{11},K_{12}}(q)\hat{\mathcal{A}}_{K_{21},K_{22}}(q)
    \cdots
    \hat{\mathcal{A}}_{K_{M1},K_{M2}}(q)
  \right\} .
\label{eq:f_q}
\end{equation}
In order to seek the critical temperature, we will investigate
$f_0 = \lim_{M\to\infty}f(0)$. We will see in fact that $f_0$ is
non-analytic at a critical point $T_{\rm c}$.

In order to simplify $\hat{\mathcal{A}}_{K_{11},K_{12}}(0)$ further, we introduce new Majorana fermions as
\begin{equation}
 c_n(q) = \frac{1}{\sqrt{2}}
  \left(\alpha_n (q) + i \beta_n (q) \right), \qquad
  c_n(q)^\dagger = \frac{1}{\sqrt{2}}
  \left(\alpha_n (q) - i \beta_n (q) \right).
\end{equation}
The transfer
matrices are rewritten in terms of these new Majorana fermions as
\begin{eqnarray}
 h_{K_{11},K_{12}}^{(1)}(q)
  &=&
  2iK_{11}\left(\alpha_2\alpha_3+\beta_2\beta_3
              \right) \nonumber \\
  && + 2iK_{12}
  \left[ \cos q \left(\alpha_4\alpha_1 + \beta_4\beta_1
                \right)
  - \sin q \left(\alpha_4 \beta_1 +\alpha_4 \beta_1 \right)
  \right] ,
\end{eqnarray}
\begin{eqnarray}
 h_g^{(2)}(q) &=& 2ig
  \left(\alpha_1\alpha_2+\alpha_3\alpha_4
  +\beta_1\beta_2+\beta_3\beta_4\right) ,
\end{eqnarray}
where we have omitted $(q)$ on $\alpha_n$ and $\beta_n$.
It is clear in the above that $\alpha_n$ and $\beta_n$ are completely decoupled and
symmetric in the limit $q\rightarrow 0$.  Hence the trace in
Eq.~(\ref{eq:f_q}) with $q\to 0$ is given by the square of the trace on the $\alpha$
subspace.

Now, focusing on $q\to 0$, we transform Majorana
operators $\alpha_n$ into the Pauli matrices $\tau_n^x$,
$\tau_n^y$ and $\tau_n^z$ as
\begin{equation}
 \alpha_1 = \frac{1}{\sqrt{2}}\tau_1^z,\quad
  \alpha_2 = \frac{1}{\sqrt{2}}\tau_1^y,\quad
  \alpha_3 = \frac{1}{\sqrt{2}}\tau_1^x \tau_2^z, \quad
  \alpha_4 = \frac{1}{\sqrt{2}}\tau_1^x \tau_2^y.
\end{equation}
Then we obtain
\begin{equation}
 2i(K_{11}\alpha_2 \alpha_3 + K_{12} \alpha_4 \alpha_1)
 = (K_{11}-K_{12} V) \tau_1^z \tau_2^z,
\end{equation}
\begin{equation}
 2ig(\alpha_1 \alpha_2 + \alpha_3 \alpha_4)
  = g (1+V) \tau_1^x,
\end{equation}
where the unitary matrix $V$ is defined as
$V= \tau_1^x\tau_2^x$.
This matrix $V$ commutes with $h_{K_{11},K_{12}}^{(1)}(0)$
and $h_g^{(2)}(0)$ and has eigenvalues $\pm 1$. Thus the transfer matrix
is decomposed into two on the subspaces with $V = \pm 1$ respectively. Therefore $f(0)$ in
Eq.~(\ref{eq:f_q}) is written as
\begin{equation}
 f(0) = -2 T \frac{1}{M}
  \ln \left({\rm tr} \prod_{i=1}^M A_i^+
   + {\rm tr} \prod_{i=1}^M A_i^- \right),\label{eq:f0_A}
\end{equation}
where $A_i^{+}$ and $A_i^{-}$ are $2\times 2$ matrices
on  the subspaces with $V=1$ and $V=-1$, respectively.
We take the basis for the $V=1$ sector as eigenstates of
$\tau_1^z\tau_2^z$, {\it i.e.},
\begin{equation}
 |+\rangle = \frac{1}{\sqrt{2}}
  \left(\left|\uparrow\uparrow\right\rangle
   + \left|\downarrow\downarrow\right\rangle \right),\qquad
 |-\rangle = \frac{1}{\sqrt{2}}
  \left(\left|\uparrow\downarrow\right\rangle
   + \left|\downarrow\uparrow\right\rangle \right).
\end{equation}
Using this set of basis, $A_i^{+}$ is written in the matrix
representation as
\begin{equation}
 A_i^{+} =
  \left(
   \begin{array}{cc}
    e^{K_{i1}-K_{i2}} & 0\\
    0 & e^{-(K_{i1}-K_{i2})}
   \end{array}
  \right)
  \left(
   \begin{array}{cc}
    \cosh 2g & \sinh 2g\\
    \sinh 2g & \cosh 2g
   \end{array}
  \right).\label{eq:matA_+}
\end{equation}
For the $V=-1$ sector, we settle the basis as
\begin{equation}
 |+\rangle = \frac{1}{\sqrt{2}}
  \left(\left|\uparrow\uparrow\right\rangle
   - \left|\downarrow\downarrow\right\rangle \right),\qquad
 |-\rangle = \frac{1}{\sqrt{2}}
  \left(\left|\uparrow\downarrow\right\rangle
   - \left|\downarrow\uparrow\right\rangle \right).
\end{equation}
Then one obtains the matrix representation of $A_i^{-}$,
\begin{equation}
 A_i^{-} =
  \left(
   \begin{array}{cc}
    e^{K_{i1}+K_{i2}} & 0\\
    0 & e^{-(K_{i1}+K_{i2})}
   \end{array}
  \right).\label{eq:matA_-}
\end{equation}

The second term in the brackets of Eq.~(\ref{eq:f0_A}), {\it i.e.},
the trace of a product of $A_i^-$ is estimated as $e^{2[K_{ij}]M}$ for large $M$, since
it is the trace of the diagonal matrices. Regarding the first term, we
define the Lyapunov exponent of a product of random matrices by
\begin{equation}
 \gamma = \lim_{M\rightarrow\infty}\frac{1}{M}
  \ln {\rm tr} \prod_{i=1}^M A_i^+.
\end{equation}
Using this, the first term is estimated as $e^{\gamma M}$. Therefore
$f_0$ is obtained finally as
\begin{equation}
 f_0 = - 2 T \max\left\{\gamma, 2[K_{ij}] \right\}.\label{eq:f0}
\end{equation}
The critical point is determined as a singular point of this quantity.

\subsection{Honeycomb lattice}

The model for the honeycomb lattice is obtained by making
$(K_{i1}, K_{i2}) = (0, K_i)$ for $i = 2,4,\cdots, M$ and
$(K_{i1}, K_{i2}) = (K_i, 0)$ for $i = 1,3,\cdots, M-1$
in the above formulas on the square lattice with $K_i \equiv J_i/T$.
Therefore one gets the following result instead of Eq.~(\ref{eq:f0}).
\begin{equation}
 f_0 = - 2 T \max
  \left\{\kappa, [K_i] \right\} ,
\label{eq:f0honeycomb}
\end{equation}
where
\begin{equation}
 \kappa = \lim_{M\rightarrow\infty} \frac{1}{M}
  \ln {\rm tr} \prod_{\mu=1}^{M/2} B_{2\mu-1}^+ B_{2\mu}^- ,
\label{eq:kappa0}
\end{equation}
\begin{equation}
 B_i^{\pm} =
  \left(
   \begin{array}{cc}
    e^{\pm K_{i}} & 0\\
    0 & e^{\mp K_{i}}
   \end{array}
  \right)
  \left(
   \begin{array}{cc}
    \cosh 2g & \sinh 2g\\
    \sinh 2g & \cosh 2g
   \end{array}
  \right).
\end{equation}
The relation between $B_i^{+}$ and $B_i^{-}$,
\begin{equation}
 B_i^{-} = U B_i^{+} U,\qquad
  U \equiv
  \left(
   \begin{array}{cc}
    0 & 1\\
    1 & 0
   \end{array}
  \right) ,
\end{equation}
yields a simpler representation
of Eq.~(\ref{eq:kappa0}) as follows.
\begin{equation}
 \kappa = \lim_{M\rightarrow\infty} \frac{1}{M}
  \ln {\rm tr} \prod_{i=1}^{M} B_i ,
\label{eq:kappa}
\end{equation}
where
\begin{equation}
 B_i \equiv B_i^{+} U =
  \left(
   \begin{array}{cc}
    e^{K_{i}} & 0\\
    0 & e^{-K_{i}}
   \end{array}
  \right)
  \left(
   \begin{array}{cc}
    \sinh 2g & \cosh 2g\\
    \cosh 2g & \sinh 2g
   \end{array}
  \right).\label{eq:matB}
\end{equation}

\section{Pure systems}
\label{sec:Pure}

For the sake of confirmation,
we visit the pure honeycomb-lattice model
in the present section, before going to the random case.  In the
pure system, $K_i$ is no longer random and, supposing $K_i = K = J_0/T$ uniformly for any $i$,
Eqs. (\ref{eq:f0honeycomb}), (\ref{eq:kappa}) and (\ref{eq:matB}) reduce to
\begin{equation}
 f_0 = - 2 T \max
  \left\{ \kappa, K \right\}
\label{eq:f0_1}
\end{equation}
and
\begin{equation}
 \kappa = \lim_{M\rightarrow\infty}\frac{1}{M}
  \ln {\rm tr} B^M
  = \ln \lambda,
\end{equation}
\begin{equation}
 B =
  \left(
   \begin{array}{cc}
    e^{K} & 0\\
    0 & e^{-K}
   \end{array}
  \right)
  \left(
   \begin{array}{cc}
    \sinh 2g & \cosh 2g\\
    \cosh 2g & \sinh 2g
   \end{array}
  \right),
\end{equation}
where $\lambda$ is the largest eigenvalues of $B$.
After a straightforward computation, one finds
\begin{equation}
 \lambda =
   \cosh K \sinh 2g +
   \sqrt{1+\cosh^2 K \sinh^2 2g} .
\end{equation}
Now, a simple algebra shows that the equation $\kappa=K$ reduces to
\begin{equation}
 \tanh K = \frac{1}{\sinh 2L}.
  \label{eq:EqTc}
\end{equation}
We define the temperature $T_{\rm c}$ as the solution of this
equation. Since we have
\begin{equation}
 f_0 =
  \left\{
   \begin{array}{@{\,}ll}
    - 2 J_0 & \mbox{for $T < T_{\rm c}$} \\
    - 2 T \ln\lambda & \mbox{for $T > T_{\rm c}$}
   \end{array}
  \right. ,
\end{equation}
$T = T_{\rm c}$ provides a singular point of $f_0$
and hence $f$.

In the special case where $L = K$, namely $J = J_0$,
Eq.~(\ref{eq:EqTc}) is further simplified into
\begin{equation}
 \cosh 2K = 2 .
\end{equation}
This is nothing but the critical point of the Ising model on the
isotropic honeycomb lattice \cite{bib:Wannier1945}.

\section{Random systems}
\label{sec:Random}

The transition points are determined by the equation
$\kappa=\left[K_i\right]$ for the honeycomb lattice and
$\gamma=2\left[K_{ij}\right]$ for the square lattice.  In order to analyze
these equations, we point out that the random matrices $A_i^{+}$ in
Eq. (\ref{eq:matA_+}) and $B_i$ in Eq. (\ref{eq:matB}) have the same
form as the transfer matrix of the one-dimensional random-field Ising
model. Therefore the Lyapunov exponent $\kappa$ and $\gamma$ can be
evaluated using techniques for the free energy of the one-dimensional
systems. In this section, we obtain the exact value of the transition
probability in the zero temperature limit and numerically calculate the
phase boundary in the finite temperature, assuming $J=J_0$.

\subsection{Zero temperature}

In the zero temperature limit, the Lyapunov exponents $\gamma$ and
$\kappa$ are regarded as ground-state energies of one-dimensional
random Ising models. In general, it is a difficult task to obtain the
exact ground-state energy of a random system. However,
the exact Lyapunov exponent in the case of $J=J_0$ can be obtained by
the method proposed in Ref. \cite{bib:Kadowaki1996}.  First we focus on
the honeycomb lattice with the $\pm J$ distribution.

We consider a one-dimensional Ising-spin system
defined by the transfer matrices $B_i$ in Eq.~(\ref{eq:matB}).
Supposing that $Z_i^{\pm}$ denote the partition functions
of this system with sites $1,2,\cdots,i$ under the fixed
boundary condition $S_i=\pm 1$ respectively, they
obey the following recursion relation
\begin{equation}
 \left(
   \begin{array}{c}
    Z_{i+1}^{+}\\
    Z_{i+1}^{-}
   \end{array}
  \right)
 = B_i
 \left(
   \begin{array}{c}
    Z_{i}^{+}\\
    Z_{i}^{-}
   \end{array}
  \right),\label{eq:recursionZ}
\end{equation}
where we assume the initial condition $Z_1^{+}=Z_1^{-}=1$.
The asymptotic behavior of $Z_i^{\pm}$ for $i\gg 1$
in the low-temperature limit $K\equiv J/T\gg 1$ should be written
in the form of
\begin{equation}
 \left(
   \begin{array}{c}
    Z_{i}^{+}\\
    Z_{i}^{-}
   \end{array}
  \right) \sim
 \left(
   \begin{array}{c}
    z^{x_i+a_i}\\
    z^{x_i}
   \end{array}
  \right),
\label{eq:asymp}
\end{equation}
where $z = e^K$ and $x_i$ and $a_i$ are random exponents
to be investigated below. We here omitted prefactors, since
they do not affect the propery of the exponent in the zero-temperature
limit.
The ground-state energy per spin of the one-dimensional system
is expressed as
\begin{equation}
 \lim_{T\rightarrow 0} \frac{\kappa}{K}
  = \lim_{M\rightarrow\infty}
  \frac{x_M+\max\{a_M,0\}}{M}
  = \lim_{M\rightarrow\infty}\frac{1}{M}
  \sum_{i=1}^{M} (x_{i+1}-x_i),
\end{equation}
where we use the fact that the exponent $a_M$ does not diverge with
$M$ as will be shown later.  The last expression implies that the
average increment of exponent $x_i$ corresponds to the Lyapunov
exponent.

When $J_i=+J$, the random matrix $B_i$ has the asymptotic form
\begin{equation}
 B_i =
  \left(
   \begin{array}{cc}
    z^{-1} & z\\
    z^{-1} & z^{-3}
   \end{array}
  \right),
\end{equation}
where we have used in Eq.~(\ref{eq:matB})
\begin{equation}
 \sinh 2g \sim z^{-2}, \qquad
  \cosh 2g \sim 1.
\end{equation}
Then the recursion relation of Eq.~(\ref{eq:recursionZ})
with Eq.~(\ref{eq:asymp}) is reduced to
\begin{eqnarray}
 x_{i+1} &=& x_i -1 + \max\{a_i, -2\}\label{eq_x_plus}\\
 a_{i+1} &=& \max\{a_i, 2\} - \max\{a_i, -2\}.
\end{eqnarray}
Similarly, when $J_i=-J$, we obtain
\begin{eqnarray}
 x_{i+1} &=& x_i +1 + \max\{a_i, -2\}\label{eq_x_minus}\\
 a_{i+1} &=& \max\{a_i, 2\} - \max\{a_i, -2\} -4.
\end{eqnarray}
Now, noting $a_1 = 0$, one can see that
possible values of $a_i$ are restricted only to $0$, $\pm 2$, and $\pm 4$.
The transition matrix which transforms $a_i$
to $a_{i+1}$ is given by
\begin{equation}
 \bordermatrix{
  & a_i=4 & a_i=2 & a_i=0 & a_i=-2 & a_i=-4 \cr
  a_{i+1}=4 & 0 & 0 & 0 & p & p \cr
  a_{i+1}=2 & 0 & 0 & p & 0 & 0 \cr
  a_{i+1}=0 & p & p & 0 & \bar{p} & \bar{p} \cr
  a_{i+1}=-2 & 0 & 0 & \bar{p} & 0 & 0 \cr
  a_{i+1}=-4 & \bar{p} & \bar{p} & 0 & 0 & 0 \cr
  },
\end{equation}
where $\bar{p} = 1 - p$. Since the Markov chain generated by this
transition matrix is clearly irreducible and aperiodic except for
non-random cases, it is ergotic and has the unique stationary
distribution corresponding to the maximum eigenvalue of one. The
distribution of $a_i$ converges in the limit $i\rightarrow \infty$ to
this stationary distribution which is explicitly given by
\begin{eqnarray}
 &&P_a({+4})=\frac{p\bar{p}}{1+\bar{p}}, \quad
  P_a({+2})=\frac{p(1-p\bar{p})}{(1+p)(1+\bar{p})}, \nonumber\\
 &&P_a(0)=\frac{1-p\bar{p}}{(1+p)(1+\bar{p})}, \nonumber\\
 &&P_a({-2})=\frac{\bar{p}(1-p\bar{p})}{(1+p)(1+\bar{p})}, \quad
  P_a({-4})=\frac{p\bar{p}}{1+p}.\nonumber
\end{eqnarray}
The increment of the exponent $x_i$ depends on $J_i$ and $\max\{a_i,-2\}$ as
shown Eqs. (\ref{eq_x_plus}) and (\ref{eq_x_minus}).  Therefore, its
average increment is computed by
\begin{equation}
 (-p+\bar{p}) + \left\{4P_a(+4) + 2P_a(+2) -2 P_a(-2) -2 P_a(-4)\right\},
\end{equation}
which yields the exact Lyapunov exponent in the zero-temperature
limit
\begin{equation}
 \lim_{T\rightarrow 0} \frac{\kappa}{K}
  = \frac{3p(1-p)}{(1+p)(2-p)}.\label{eq:kappa_zero}
\end{equation}

Finally, noting $[K_i] = (2p - 1)K$ for the $\pm J$
model, we obtain
the exact critical probability $p_c$ in the
zero-temperature limit as
\begin{equation}
 p_{\rm c} = 1-2 \sin\frac{\pi}{18}
  = 0.65270364\cdots.
\label{eq:p_c_hex}
\end{equation}

A similar calculation for the square lattice yields
\begin{equation}
 \lim_{T\rightarrow 0} \frac{\gamma}{K}
  = 2 p(1-p),
\end{equation}
and the critical probability
\begin{equation}
 p_{\rm c} = \frac{\sqrt{5}-1}{2} = 0.61803399\cdots ,
\end{equation}
which is equal to the inverse of the golden ratio.

In the diluted models, the Lyapunov exponents are obtained as
\begin{equation}
 \lim_{T\rightarrow 0} \frac{\kappa}{K}
  = \frac{p(1-p)}{3-p},\qquad
 \lim_{T\rightarrow 0} \frac{\gamma}{K}
  = \frac{2p(1-p)}{3}.
\end{equation}
The critical probability in the zero-temperature limit is $p_{\rm
c}=0$ both for the honeycomb and square lattices.
This result as well as the known fact that the pure system
($p = 1$) has the ferromagnetic ground state lead to the conclusion
that the zero-temperature phase of the diluted model for $p>0$ is
ferromagnetic.
This conclusion is naturally understood since this model is always
percolated except $p=0$.

\subsection{Finite temperatures}

So far some methods have been developed to numerically compute the
Lyapunov exponent of a product of random matrices. Mainieri proposed the
cycle expansion method whose convergence is exponentially fast in the
cycle length \cite{bib:Mainieri1992}. Bai improved this method using the
evolution operator approach \cite{bib:Bai2007}.  We emply this improved
cycle expansion (ICE) method and the Monte Carlo method.  The ICE method
is summarized briefly in Appendix.
Although we still consider $J=J_0$ for comparison with the zero-temperature
results, numerical methods used in this section can be applied to $J\neq
J_0$.

\begin{figure}
 \begin{center}
  \includegraphics[scale=1.0]{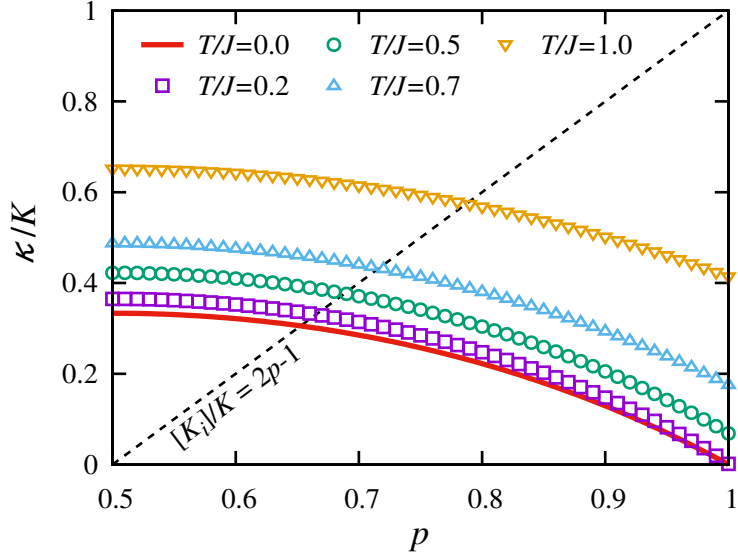}
 \end{center}

 \caption{$p$ dependence of the Lyapunov exponent for the $\pm J$ model
 on the honeycomb lattice. The dashed line indicates $[K_i]/K$.  The
 critical point corresponds to an intersection point between the curve
 of the Lyapunov exponent and the dashed line.}
 \label{fig:Lyapunov_hex_pmJ}
\end{figure}

\begin{figure}
 \begin{center}
  \includegraphics[scale=1.0]{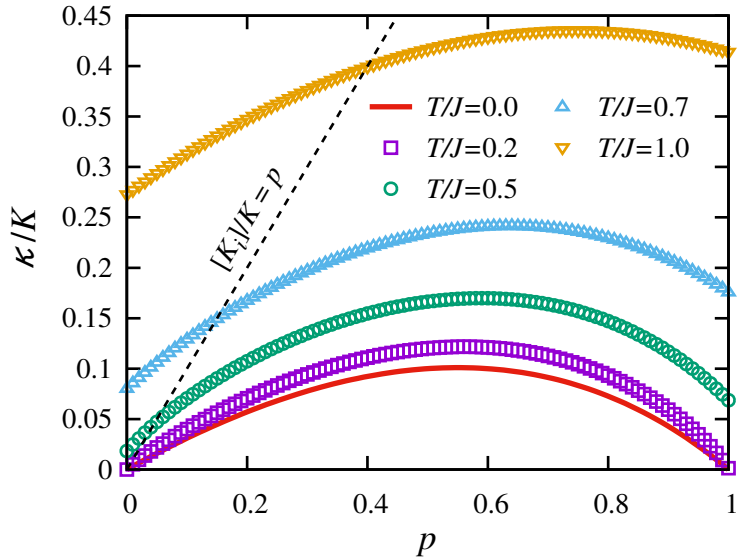}
 \end{center}
 \caption{$p$ dependence of the Lyapunov exponent
for the diluted model on the honeycomb lattice.}
 \label{fig:Lyapunov_hex_dilute}
\end{figure}

For the honeycomb-lattice model with $T/J \geq 0.7$,
we used the ICE with the cycle length $n=20$, which
produces more than $15$ reliable digits. For lower temperatures, however,
the ICE needs a still longer cycle length to attain the convergence.
This cost was complemented with the Monte Carlo method for $T/J < 0.7$,
where Lyapunov exponents were obtained by averaging over $1000$
configurations of a product of
$10^5$ matrices with the standard error of $\kappa/K$ less than
$6.0\times 10^{-5}$.

As for the square lattice, since the random matrix $A_i^{+}$ has three
possible choices, the ICE needs more computational cost as well as
longer cycle length for convergence than the honeycomb lattice.
Therefore we used the ICE with the cycle length $n=15$ for
$T/J\geq 1.0$ and the Monte Carlo method for $T/J<1.0$. Results by ICE
have the same accuracy as the honeycomb lattice, while the standard
errors of $\gamma/K$ by the Monte Carlo method are less than
$9.0\times 10^{-5}$.

The $p$ dependence of the Lyapunov exponent for the $\pm J$ model on
the honeycomb lattice is shown in Fig. \ref{fig:Lyapunov_hex_pmJ}. The solid curve
indicates the
exact result (\ref{eq:kappa_zero}) in the zero-temperature limit.  The
critical points are determined by the intersection point
with the dashed line ($[K_i]/K=2p-1$).
The zero-temperature critical point is given exactly by Eq.~(\ref{eq:p_c_hex}).
In Fig. \ref{fig:Lyapunov_hex_dilute} we show
the results for the diluted model. It is clear that the
critical probability in the zero-temperature limit is equal to zero as
mentioned before.

\begin{figure}
 \begin{center}
  \includegraphics[scale=1.0]{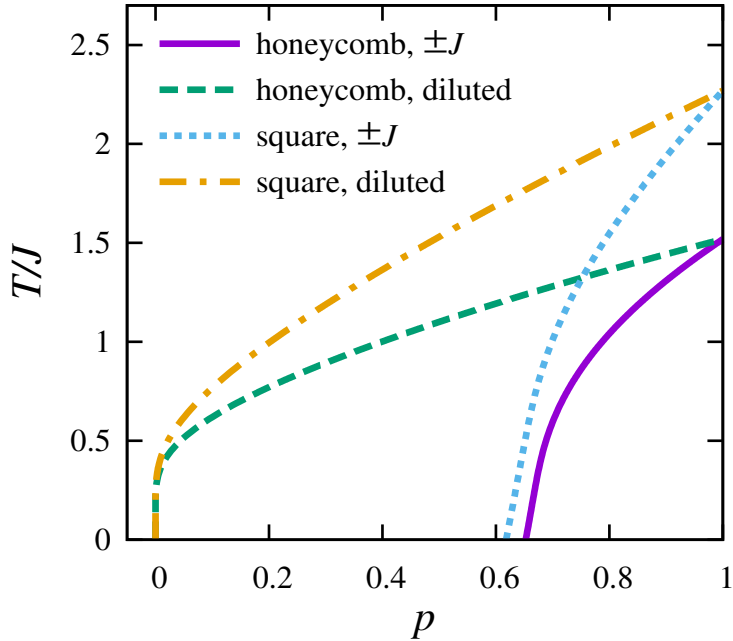}
 \end{center}
 \caption{Critical lines in the $p-T$ plane
of the $\pm J$ models and the diluted models on
the honeycomb and squara lattices.}
 \label{fig:phase}
\end{figure}

The critical line in the $p-T$ plane
were obtained by solving the equation $\kappa = [K_i]$ or
$\gamma = 2[K_{ij}]$. For the temperature
region where the ICE is valid, we estimated the critical
temperature using the bisection method for given $p$'s.
On the other hand, for lower temperatures,
we obtained the critical probability for
given $T$'s using the Monte Carlo method and the bisection method.
The results are shown in Fig. \ref{fig:phase}.
Although one cannot say anything about the property of the
phases away from the critical line, it has been shown exactly
that the critical temperature at $p=1$ is the transition
point between the high-temperature paramagnetic phase
and the low-temperature ferromagnetic phase.
This fact as well as the fact that the critical line obtained here
is continuously connected with the $p=1$ transition point
imply that the critical line is in fact the transition line
separating the paramagnetic and ferromagnetic phases.
It is worth noting that the $\pm J$ model has the
ferromagntic-antiferromagnetic symmetry with respect to
$p = 1/2$, as mentioned in Sect.~\ref{sec:Model}.
Therefore, we deduce that there is a striped
antiferromagnetic phase in the small $p$ region, though
it is not shown in Fig.~\ref{fig:phase}.
In addition, the study on the Shanker-Murthy model
suggests that
the Griffiths phase may exist between the ferromagnetic and
antiferromagnetic phases and below the critical temperature of the pure
system \cite{bib:Griffiths1969,bib:Randeria1985}.

We comment that Hoever \cite{bib:Hoever1982} has numerically studied the
critical line of the $\pm J$
model on the square lattice. Our result on the same model is
in perfect agreement with it.

\begin{figure}
 \begin{center}
  \includegraphics[scale=1.0]{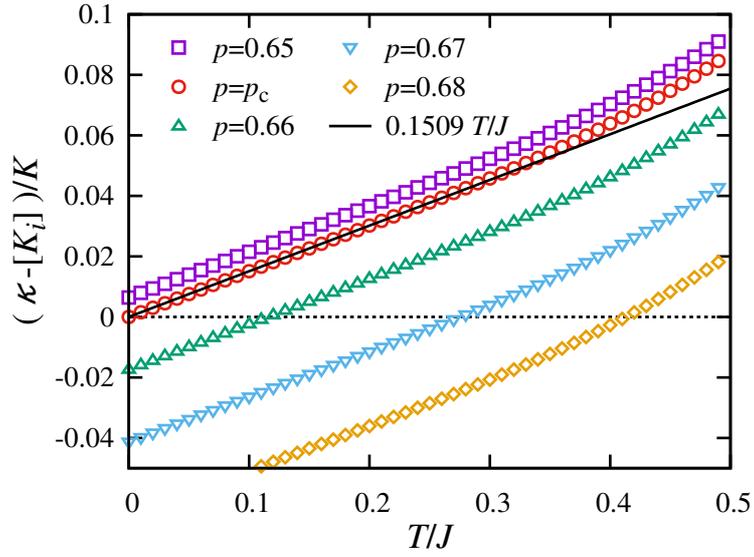}
 \end{center}
 \caption{Low-temperature behavior of the Lyapunov exponent for the $\pm
 J$ model on the honeycomb lattice. The solid line is obtained by
 fitting the data at $p=p_c$ for $0\leq T/J < 0.2$. The intersection
 with the horizontal dashed line indicates the transition temperature.}
 \label{fig:Lyapunov_hex_pmJ_vs_t}
\end{figure}

It should be pointed that the critical lines of
the $\pm J$ model on both lattices are not vertical near
zero temperature on the $p-T$ plane.
As shown in Fig. \ref{fig:Lyapunov_hex_pmJ_vs_t}, the
Lyapunov exponent $\kappa$ devided by $K$ at low temperature seems to have a form
\begin{equation}
 \kappa/K \simeq e_0(p) + c T,
\end{equation}
where $e_0(p)$ denotes the zero-temperature limit of $\kappa/K$
which is exactly given by
Eq. (\ref{eq:kappa_zero}). The coefficient $c$ is estimated as
$c=0.1509$ by fitting the data at $p=p_c$ for $0\leq T/J < 0.2$.
This observation implies that the critical temperature for $T/J \ll 1$ is
behaves as
\begin{equation}
 T_c(p) \simeq \frac{1}{c}\left\{ (2p-1)-e_0(p) \right\}
  \simeq 15.70 \times (p-p_c),
\end{equation}
which is consistent with our numerical result (Fig. \ref{fig:phase}). On
the square lattice, we obtain $T_c(p)\simeq 14.64 \times (p-p_c)$.

These behaviors of the critical temperature are quite
different from the Shankar-Murthy model, where
the constraint $K_{i1}=K_{i2}$ leads to the exact
Lyapunov exponent $\gamma=2g$ regardless of $p$.
Then an essential singularity of $g \sim e^{-2J/T}$ at $T=0$ causes the
vertical growth of the critical temperature at $p_c=1/2$.

\begin{figure}
 \begin{center}
  \includegraphics[scale=1.0]{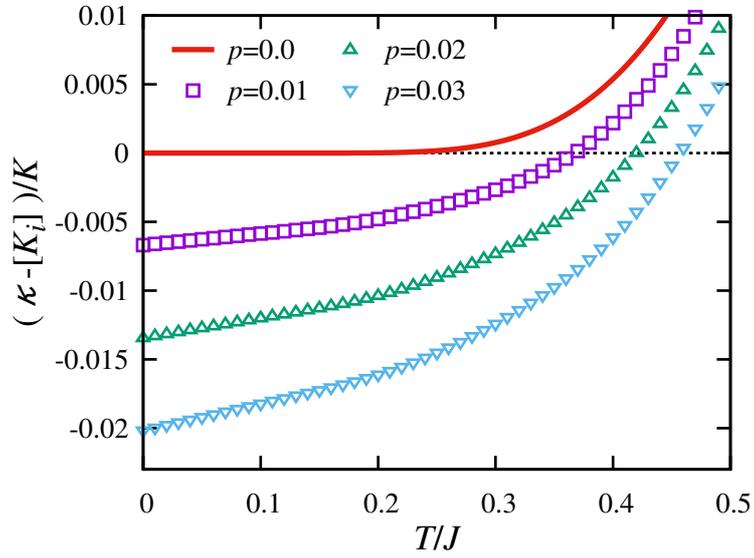}
 \end{center}
 \caption{Low-temperature behavior of the Lyapunov exponent for the
 diluted  model on the honeycomb lattice.}
 \label{fig:Lyapunov_hex_dilute_vs_t}
\end{figure}

In contrast to the $\pm J$ models,
the diluted model has the vertical phase boundary in the same way as the
Shankar-Murthy model. The Lyapunov
exponent at $p=0$ is exactly calculated as $\kappa=2g$
for the honeycomb lattice, which is
essentially singular with respect to $T$.  The low-temperature behavior
of the Lyapunov exponent is shown in Fig. \ref{fig:Lyapunov_hex_dilute_vs_t}.

\section{Conclusion}
\label{sec:Concl}

In this paper, we discussed the two-dimensional Ising model with
striped randomness on the honeycomb lattice and on the square
lattice. We simplified the transfer matrices by using the Majorana
fermion operators and obtained the $2\times 2$ matrix representation in
the long-wavelength limit. The Lyapunov exponents can be calculated by
regarding these decomposed matrices as the transfer matrix of the
one-dimensional random-field Ising model. We obtained its exact solution
in the zero-temperature limit and determined the exact value
for the critical probability $p_c$ of ferromagnetic bonds.
The critical line on the probability-temperature
($p-T$) plane was
also calculated with highly accurate numerical methods.

We focus only on the zero wave-number
limit to analyze the critical
point. However the wave-number dependence of the free energy is necessary to
analyze critical phenomena and to determine the universality class,
which remains as one of the future issues.

\appendix
\section{Cycle expansion of the Lyapunov exponent}

In this appendix, we briefly summarize the improved cycle expansion method
as a numerical method to compute the Lyapunov exponent.

The Lyapunov exponent of a product of random matrices is defined as
\begin{equation}
 \gamma = \lim_{n\rightarrow\infty} \frac{1}{n}
  \left\langle
   \ln \left\| M_1 M_2 \cdots M_n \right\|
  \right\rangle
\end{equation}
where $M_i$ is a random matrix and the angular brackets denote the
ensemble average. It is notable that the Lyapunov exponent is
independent of a choice of the matrix norm $\| \cdot \|$ as long as
equivalent norm used.  The maximum eigenvalue $\mu_0(M)$ is the most
convenient norm to derive the cycle expansion because it is invariant under
cyclic rotation $\mu_0(M_iM_j)=\mu_0(M_jM_i)$ and satisfies
$\mu_0(M^2)=\mu_0(M)^2$.

Assume that the random matrix $M_i$ has $m$ possible
choices to be $M_i=A_s$
with probability $p_s$ $(s=1,2,\cdots,m)$.  With a string of $n$ choices
$S\equiv s_1s_2\cdots s_n$, the product of $n$ random matrices
$A_{s_1}A_{s_2}\cdots A_{s_n}$ is denoted by $A_S$ and its probability
is by $P(S) \equiv \prod_{i=1}^n p_{s_i}$.  With these notations,
the Lyapunov exponent is expressed as the large-$n$ limit of the
following quantity
\begin{equation}
 \gamma_n \equiv \frac{1}{n}\sum_{|S|=n}
  P(S) \ln \mu_0(A_S),\label{eq:def_gamma_n}
\end{equation}
where $|S|$ is the length of a string $S$.

Owing to the properties of $\mu_0(A)$, the sum in
Eq. (\ref{eq:def_gamma_n}) is decomposed into the sum over the set $C_p$
of all possible primitive cycles, where a cycle is
said primitive if it is not a
repeat of a smaller length cycle.  Two cycles are equivalent if they
differ only by a cyclic rotation.  For example, $s_1s_2s_1s_2\not\in
C_p$ because it is a repeat of $s_1s_2$, and if $s_1s_2s_3\in C_p$,
$s_2s_3s_1\not\in C_p$.  As a result, the cycle expansion of the
Lyapunov exponent is obtained as
\begin{equation}
 \gamma_n = \sum_{S\in C_p}\sum_{r=1}^{n} \delta_{r|S|,n}
  P(S)^r \ln \mu_0(A_S).
\end{equation}
This formula was also derived from the expansion of the thermodynamic
zeta function and its exponential convergence with the cycle length was
observed \cite{bib:Mainieri1992}.

Bai proposed an accelerated algorithm for the cycle expansion based on
the evolution operator approach \cite{bib:Bai2007}.
Bai numerically showed super-exponential convergence of the
weighted average of the cycle expansion,
\begin{equation}
 \tilde{\gamma}_n =
  \frac{\sum_{k=0}^{n-1} w_k\gamma_{n-k}}{\sum_{k=0}^{n-1} w_k},
\end{equation}
in contrast to exponential convergence of the cycle expansion method.
The averaging weight $w_k$ is determined to eliminate all exponential
converging terms by using the evolution operator approach.
In the case of $2\times 2$ matrices with positive elements, $w_k$ is
given by
\begin{equation}
 w_k = -\frac{1}{k} \sum_{j=1}^{k} c_j w_{k-j},
\end{equation}
\begin{equation}
 c_n = \sum_{S\in C_p}\sum_{r=1}^{n} \delta_{r|S|,n}|S|
  \frac{\left[P(S) g(A_S)\right]^r}{1-g(A_S)^r},
\end{equation}
where $g(A)$ denotes the ratio of the second eigenvalue of $A$ to
the first one $\mu_0(A)$.

\subsubsection*{acknowledgements}
SM would like to thank T. Hamasaki and H. Nishimori for helpful
advices.
The work of SS was supported by the JSPS (grant No. 26400402).


\begin{thebibliography}{}
 \bibitem{bib:Aharony1980}
         Aharony, A., Stephen, M.J.:
         Duality relations and the replica method for Ising models.
         \newblock J. Phys. C: Solid State Phys. \textbf{13}, L407--L414 (1980)

\bibitem{bib:Bai2007}
        Bai, Z.Q.:
        On the cycle expansion for the Lyapunov exponent of a product of random matrices.
        \newblock J. Phys. A: Math. Theor. \textbf{40}, 8315--8328 (2007)

\bibitem{bib:Chaloupka2010}
        Chaloupka, J., Jackeli, G., Khaliullin, G.:
        Kitaev-Heisenberg model on a honeycomb lattice: Possible exotic phases in iridium oxides
        ${A}_{2}{\mathrm{IrO}}_{3}$.
        \newblock Phys. Rev. Lett. \textbf{105}, 027204 (2010)

\bibitem{bib:Chaloupka2013}
        Chaloupka, J., Jackeli, G., Khaliullin, G.:
        Zigzag magnetic order in the iridium oxide ${\mathrm{Na}}_{2}{\mathrm{IrO}}_{3}$.
        \newblock Phys. Rev. Lett. \textbf{110}, 097204 (2013)

\bibitem{bib:Choi2012}
        Choi, S.K., Coldea, R., Kolmogorov, A.N., Lancaster, T., Mazin,
        I.I., Blundell, S.J., Radaelli, P.G., Singh, Y., Gegenwart, P.,
        Choi, K.R., Cheong, S.W., Baker, P.J., Stock, C., Taylor, J.:
        Spin waves and revised crystal structure of honeycomb iridate
        ${\mathrm{Na}}_{2}{\mathrm{IrO}}_{3}$.
        \newblock Phys. Rev. Lett. \textbf{108}, 127204 (2012)

\bibitem{bib:Domany1978}
        Domany, E.:
        Criticality and crossover in the bond-diluted random Ising model.
        \newblock J. Phys. C: Solid State Phys. \textbf{11}, L337--L342 (1978)

\bibitem{bib:Fisch1978}
        Fisch, R.:
        Critical temperature for two-dimensional Ising ferromagnets with
        quenched bond disorder.
        \newblock J. Stat. Phys. \textbf{18}, 111--114 (1978)

\bibitem{bib:Griffiths1969}
        Griffiths, R.B.:
        Nonanalytic behavior above the critical point in a random Ising ferromagnet.
        \newblock Phys. Rev. Lett. \textbf{23}, 17--19 (1969)

\bibitem{bib:Hamm1977}
        Hamm, J.R.:
        Regularly spaced blocks of impurities in the Ising model:
        Critical temperature and specific heat.
        \newblock Phys. Rev. B \textbf{15}, 5391--5411 (1977)

\bibitem{bib:Hoever1982}
        Hoever, P.:
        Generalized transfer formalism and application to random Ising models.
        \newblock Z. Phys. B Condensed Matter \textbf{48}, 137--148 (1982)

\bibitem{bib:Houtappel1950}
        Houtappel, R.:
        Order-disorder in hexagonal lattices.
        \newblock Physica \textbf{16}, 425--455 (1950)

\bibitem{bib:Igloi1995}
        Igl\'{o}i, F., Lajk\'{o}, P.:
        On the critical temperature of non-periodic Ising models on hexagonal lattices.
        \newblock Z. Phys. B Condensed Matter \textbf{99}, 281--283 (1995)

\bibitem{bib:Jackeli2009}
        Jackeli, G., Khaliullin, G.:
        Mott insulators in the strong spin-orbit coupling limit:
        From Heisenberg to a quantum compass and Kitaev models.
        \newblock Phys. Rev. Lett. \textbf{102}, 017205 (2009)

\bibitem{bib:Kadowaki1996}
        Kadowaki, T., Nonomura, Y., Nishimori, H.:
        Exact ground-state energy of the Ising spin glass on strips.
        \newblock J. Phys. Soc. Jpn. \textbf{65}, 1609--1616 (1996)

\bibitem{bib:Kardar1982}
        Kardar, M., Berker, A.N.:
        Exact criticality condition for randomly layered Ising models
        with competing interactions on a square lattice.
        \newblock Phys. Rev. B \textbf{26}, 219--225 (1982)

\bibitem{bib:Kaufman1949}
        Kaufman, B.:
        Crystal statistics. II. Partition function evaluated by spinor analysis.
        \newblock Phys. Rev. \textbf{76}, 1232--1243 (1949)

\bibitem{bib:Kitaev2006}
        Kitaev, A.:
        Anyons in an exactly solved model and beyond.
        \newblock Ann. Phys. \textbf{321}, 2--111 (2006)

\bibitem{bib:Kramers1941}
        Kramers, H.A., Wannier, G.H.:
        Statistics of the two-dimensional ferromagnet. Part I.
        \newblock Phys. Rev. \textbf{60}, 252--262 (1941)

\bibitem{bib:Mainieri1992}
        Mainieri, R.:
        Zeta function for the Lyapunov exponent of a product of random matrices.
        \newblock Phys. Rev. Lett. \textbf{68}, 1965--1968 (1992)

\bibitem{bib:McCoy1968}
        McCoy, B.M., Wu, T.T.:
        Theory of a two-dimensional Ising model with random impurities. I. Thermodynamics.
        \newblock Phys. Rev. \textbf{176}, 631--643 (1968)

\bibitem{bib:Onsager1944}
        Onsager, L.:
        Crystal statistics. I. A two-dimensional model with an order-disorder transition.
        \newblock Phys. Rev. \textbf{65}, 117--149 (1944)

\bibitem{bib:Randeria1985}
        Randeria, M., Sethna, J.P., Palmer, R.G.:
        Low-frequency relaxation in Ising spin-glasses.
        \newblock Phys. Rev. Lett. \textbf{54}, 1321--1324 (1985)

\bibitem{bib:Reuther2011}
        Reuther, J., Thomale, R., Trebst, S.:
        Finite-temperature phase diagram of the Heisenberg-Kitaev model.
        \newblock Phys. Rev. B \textbf{84}, 100406 (2011)

\bibitem{bib:Schwartz1979}
        Schwartz, M.:
        Dual relations for quenched random systems.
        \newblock Phys. Lett. A \textbf{75}, 102--104 (1979)

\bibitem{bib:Shankar1987}
        Shankar, R., Murthy, G.:
        Nearest-neighbor frustrated random-bond model in $d=2$: Some exact results.
        \newblock Phys. Rev. B \textbf{36}, 536--545 (1987)

\bibitem{bib:Singh2012}
        Singh, Y., Manni, S., Reuther, J., Berlijn, T., Thomale, R., Ku, W., Trebst,
        S., Gegenwart, P.:
        Relevance of the Heisenberg-Kitaev model for the honeycomb
        lattice iridates ${A}_{2}{\mathrm{IrO}}_{3}$.
        \newblock Phys. Rev. Lett. \textbf{108}, 127203 (2012)

\bibitem{bib:Wannier1945}
        Wannier, G.H.:
        The statistical problem in cooperative phenomena.
        \newblock Rev. Mod. Phys. \textbf{17}, 50--60 (1945)

\bibitem{bib:Wannier1950}
        Wannier, G.H.:
        Antiferromagnetism. The triangular Ising net.
        \newblock Phys. Rev. \textbf{79}, 357--364 (1950)

\bibitem{bib:Wolff1981}
        Wolff, W.F., Hoever, P., Zittartz, J.:
        Layered inhomogeneous Ising models with frustration on a square lattice.
        \newblock Z. Phys. B Condensed Matter \textbf{42}, 259--264 (1981)

\bibitem{bib:Ye2012}
        Ye, F., Chi, S., Cao, H., Chakoumakos, B.C., Fernandez-Baca, J.A., Custelcean,
        R., Qi, T.F., Korneta, O.B., Cao, G.:
        Direct evidence of a zigzag spin-chain structure in the honeycomb
        lattice: A neutron and X-ray diffraction investigation of
        single-crystal ${\mathrm{Na}}_{2}{\mathrm{IrO}}_{3}$.
        \newblock Phys. Rev. B \textbf{85}, 180403 (2012)

\end{thebibliography}
\end{document}